# Accelerating Combinatorial Electrocatalyst Discovery with Bayesian Optimization: A Case Study in the Quaternary System Ni-Pd-Pt-Ru for the Oxygen Evolution Reaction

F. Thelen[1], R. Zehl[1], R. Zerdoumi[2], J. L. Bürgel[1], L. Banko[1], W. Schuhmann[2], A. Ludwig[1],*

[1]Chair for Materials Discovery and Interfaces, Institute for Materials, Faculty of Mechanical Engineering, Ruhr University Bochum, Universitätsstraße 150, 44801 Bochum, Germany

[2]Analytical Chemistry– Center for Electrochemical Sciences (CES), Faculty of Chemistry and Biochemistry, Ruhr University Bochum, Universitätsstraße 150, 44801 Bochum, Germany

Underlined authors contributed equally.

*corresponding author: alfred.ludwig@rub.de

**Abstract**

The discovery of high-performance electrocatalysts is crucial for advancing sustainable energy technologies. Compositionally complex solid solutions comprising multiple metals offer promising catalytic properties, yet their exploration is challenging due to the combinatorial explosion of possible compositions. In this work, we combine combinatorial sputtering of thin-film materials libraries and their high-throughput characterization with Bayesian optimization to efficiently explore the quaternary composition space Ni-Pd-Pt-Ru for the oxygen evolution reaction in alkaline media. Using this method, the global activity optimum of pure Ru was identified after covering less than 20% of the complete composition space with six materials libraries. Six additional libraries were fabricated to validate the activity trend. The resulting dataset is used to formulate general guidelines for the efficient composition space exploration using combinatorial synthesis paired with Bayesian optimization.

**Introduction**

The search for high-performance electrocatalysts is one of the key challenges in sustainable energy research. Many of the best-known catalysts rely on scarce and expensive elements, limiting their large-scale application. Compositionally complex solid solutions, which are quaternary, quinary or higher-order systems with multiple elements, frequently present in nearly-equiatomic ratios, offer a promising pathway to overcome these limitations by enabling tunable electronic structures and adsorption properties [1]. However, even when considering a resolution of 5 at.%, a quaternary system already contains 1,771 possible compositions, while a quinary system expands to 15,504 compositions. Therefore, identifying optimal compositions within the vast composition space of multinary systems is challenging, making traditional experimental approaches impractical [2].

Combinatorial materials synthesis techniques, such as magnetron co-sputtering [3], can address this challenge by enabling the fabrication of large numbers of well-defined compositions in parallel. These methods produce materials libraries, which consist of thin-film continuous compositional gradients or discrete compositions on a single substrate. Paired with high-throughput characterization (screening) methods, which allow the automated simultaneous or serial investigation of hundreds of compositions, the combinatorial approach can improve the efficiency significantly compared to one-at-a-time experiments [4]. Figure 1a summarizes the time effort required for the synthesis and characterization of a single materials library for electrochemical applications with 342 pre-defined measurement areas. While synthesis is relatively fast, characterization is often the most time-consuming step, even when limited to essential techniques. Energy-dispersive X-ray spectroscopy (EDX), used to determine the thin-film (volume) composition, is one of the shortest characterization steps. X-ray diffraction (XRD) follows, providing insights into the material's crystal structure and phase constitution. The electrochemical screening, performed using a scanning droplet cell (SDC) setup [5], requires the longest time. The overall process, excluding analysis time, typically spans about two days. While this high-throughput approach significantly accelerates materials discovery and hundreds of compositions can be fabricated in parallel, it is still constrained by the combinatorial explosion. In quaternary or higher-order systems, a single materials library can only cover a limited fraction of the composition space: a quaternary library typically spans about 5%, while the coverage of a quinary system can be less than 1% [6]. Consequently, exploring the composition space requires the iterative fabrication of multiple materials libraries, each covering a distinct compositional region. The resulting exploration cycle is summarized in Figure 1b. By varying the cathode powers in each deposition, materials libraries can be fabricated with different compositional ranges: each library corresponds to a hyperbolic paraboloid plane in the multidimensional composition space.

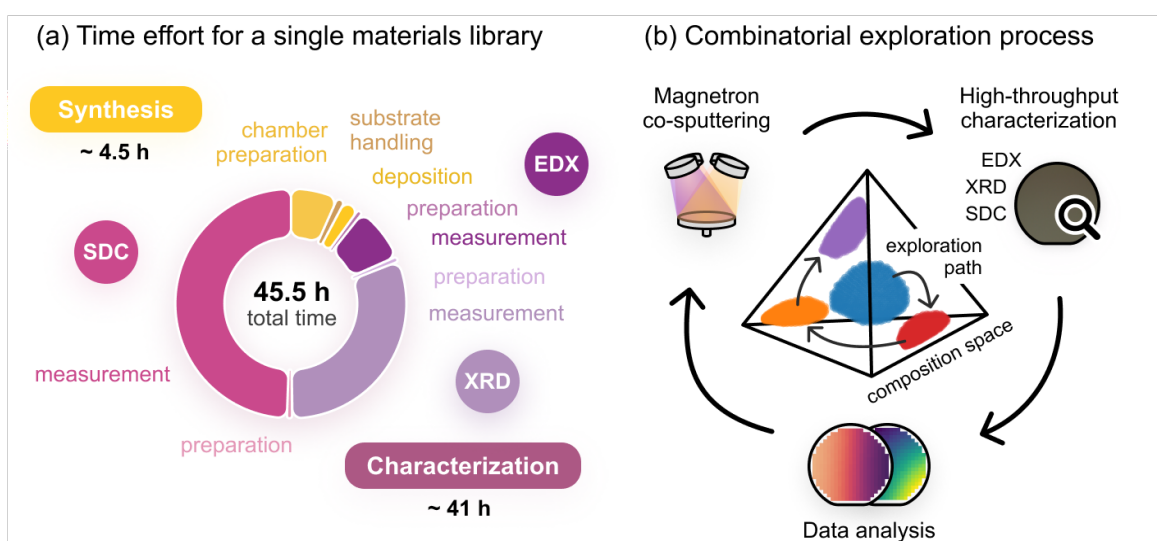

**Figure 1:** (a) Estimated time required for the synthesis and characterization of a single materials library for electrochemical applications. While deposition is fast, characterization takes significantly longer. (b) Schematic of the iterative exploration cycle, where multiple libraries are fabricated to cover the composition space. The next library is fabricated based on the already acquired data.

Given these constraints, the fabrication of each new materials library must be guided by the data already acquired from previous experiments. However, navigating multidimensional composition spaces—where both compositional trends as well as electrochemical activity distributions must be considered—is a significant challenge for human intuition alone. Bayesian optimization can be a powerful framework to assist researchers in this task as it closely resembles the exploration cycle of fabrication, characterization and analysis. With this approach, a machine learning model can predict the unknown regions of the composition space based on the previously acquired data in this system [7]. The model selects new compositions for fabrication using an acquisition function, such as the expected improvement function. This determines whether the next experiment should focus on exploring less-characterized regions of the composition space or exploiting areas that already show promising activity [8,9]. Mostly, a Gaussian process is used for the underlying machine learning model due to its flexibility and ability of uncertainty quantification independent from the actual observations. Additionally, Gaussian processes are well suited for small and high-dimensional datasets [8,10].

Due to its similarity to the materials exploration cycle, Bayesian optimization has been frequently applied in materials discovery [11–16]. Pairing it with combinatorial synthesis and high-throughput characterization techniques offers a key advantage: instead of improving the model one composition at a time, entire compositional gradients can be added to the training data in a single iteration. This enhances the efficiency of data collection for materials exploration, accelerating the data-driven search for new catalysts.

**Results**

The combinatorial composition space exploration was performed in the quaternary system Ni-Pd-Pt-Ru to efficiently identify the most electrocatalytically active composition for the oxygen evolution reaction (OER) in alkaline media. The constituents were selected based on a recent exploration in the Ag-Pd-Pt-Ru system [17]. By the substitution of Ag with Ni, a more earth-abundant and cost-effective alternative element was chosen, while keeping the same elemental face-centered cubic crystal structure of Ag. The resulting system provides a suitable framework to evaluate the Bayesian optimization driven composition space exploration, as the catalytic activity is expected to increase with Ru content, due to the well-established role of Ru and its oxides as highly efficient OER catalysts [18]. The materials libraries were fabricated in a magnetron co-sputtering system with four cathodes. The chemical compositions of the libraries were characterized by EDX and the electrocatalytic activity with SDC measurements. The Bayesian optimization loop was implemented with a standard Gaussian process and the expected improvement acquisition function. The model was supplied with the measured compositions as input training data. Since the Gaussian process cannot handle voltammetric data—such as linear sweep voltammograms (LSVs) obtained by SDC measurements—out of the box, the current densities at a potential of 1.7 V vs. the reversible hydrogen electrode (RHE) were extracted from the LSVs. These values were used as a one-dimensional activity measure and added to the model for the output training data. A detailed description of the experimental methods as well as the Bayesian optimization model can be found in the supporting information.

The expected improvement acquisition function allows to balance the uncertainty of the model and the already found activity maximum. However, during the first few learning iterations, the algorithm prioritizes exploration of the corners of the composition space due to the overall large search space and the resulting high uncertainty of the Gaussian process in these regions. Therefore, initializing the Bayesian optimization with $n + 1$ materials libraries covering the center as well as the corners of the compositional space reduces the experimental time. In order to assess the coverage of the libraries in the composition space, a coverage measure was developed based on a k-nearest neighbor algorithm. More details are provided in the supporting information.

The compositions of the five initial materials libraries are visualized in the quaternary composition space in Figure 2. The libraries cover 18.6% of the quaternary composition space. The equiatomic library (ML1) achieves the highest compositional coverage with 5.7%, while the Pt-rich library (ML4) shows the smallest coverage of 2%. Due to the simultaneous operation of all four cathodes during co-sputtering, compositions at the extreme boundaries of the quaternary space are inaccessible.

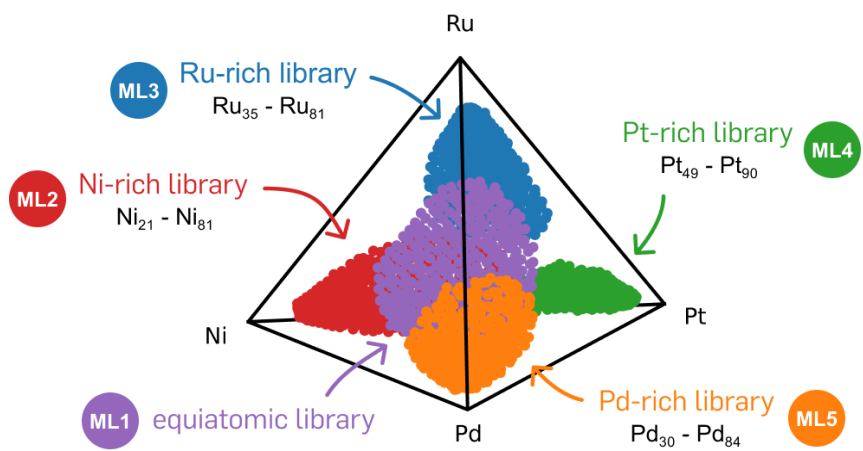

**Figure 2:** Compositions of the five initial materials libraries plotted in the quaternary composition space. The equiatomic library (purple), the Ru-rich (blue) as well as the Pd-rich library (orange) feature overlapping compositions. In this 2D representation of the tetrahedron, the Ni-rich and Pt-rich libraries appear to overlap with the others, although this is not the case when viewed in 3D. The subscript values indicate the content of each dominating element in the materials libraries in atomic percent. An animated version of this plot can be found in the supporting information.

Figure 3a exemplarily shows the LSVs obtained from SDC measurements of the equiatomic materials library, color-coded by their current density at a potential of 1.7 V vs. RHE and the resulting activity distribution on ML1. The activity in this library increases with higher Ru-contents. Figure 3b combines compositional data with measured activities, forming the training dataset for the Bayesian optimization algorithm. The highest measured activity of 1.65 mA/cm$^2$ was found in the Ru-rich library within the high Ru content region. Towards the equiatomic library, the activity decreases gradually, forming a smooth activity gradient.

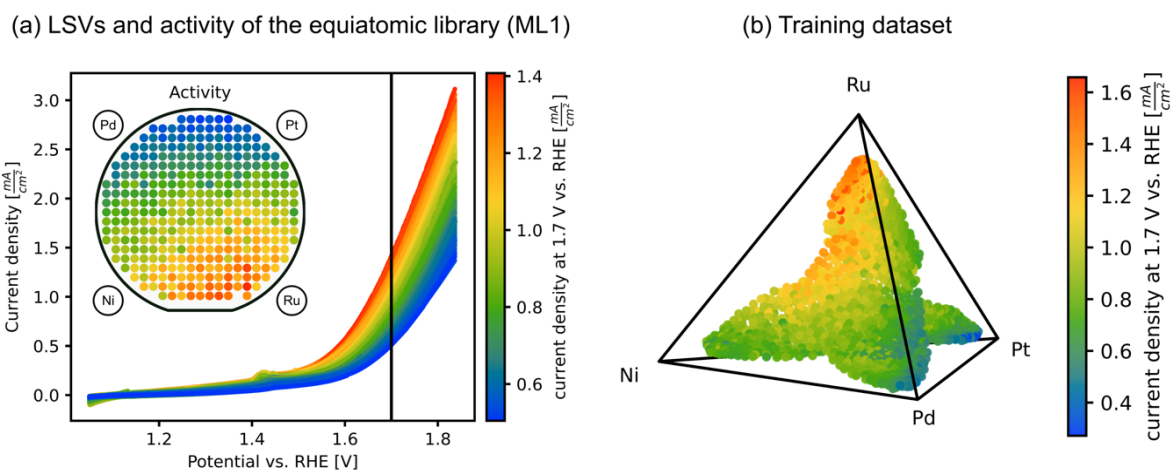

**Figure 3:** (a) LSVs obtained from the SDC measurements of the equiatomic materials library (ML1). The activity is determined by the current density at 1.7 V. The sputter target positions in the fabrication process were identical across all depositions. The activity increases with higher Ru contents. By combining the activity with the compositions of all five materials libraries, the training dataset for the Bayesian optimization algorithm, visualized in (b), is created.

After training, the Gaussian process model predicts the activity across the remaining composition space, as illustrated in Figure 4a. Due to the model's unconstrained output space, negative current densities are occasionally predicted, notably in the compositional region near Pd$_2$Ru. Consistent with the training dataset, the model predicts high activity for compositions with higher Ru content. While the incorporation of Pd and Pt shows no positive impact on activity, the model suggests that Ni contents ranging from 5 to 20 at.% may enhance performance. Since the model also smooths the training data, it effectively removes experimental noise, which can result in predicted activity values lower than the highest measured activity. Together with the uncertainty of the model, the prediction is used to compute the expected improvement acquisition function on the composition space, indicating in which compositional range to sample next. This is visualized in Figure 4b. A high expected improvement was found in the ternary Ni-Pd-Ru region with high Ru content. The composition showing the highest expected improvement is $Ni_{15}Ru_{85}$.

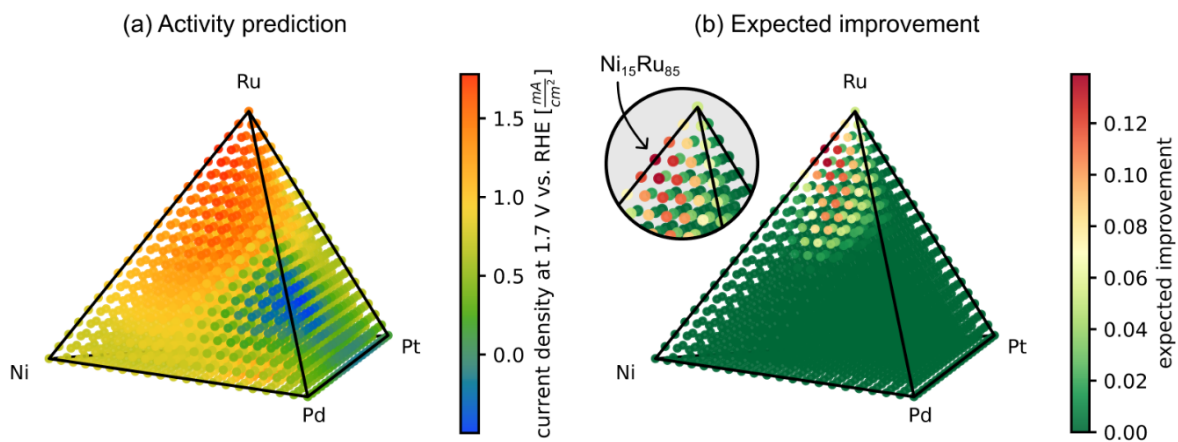

**Figure 4:** (a) Activities predicted by the Gaussian process based on the five initial materials libraries (ML1-5). The activity trend points towards compositions with a high Ru-content and with higher contents of Ni than Pd or Pt. (b) Expected improvement calculated from the activity prediction and the model's uncertainty in the composition space sampled at 5 at.% steps. The composition with the highest expected improvement is $Ni_{15}Ru_{85}$.

Based on the above findings and to improve the model, the next materials library (ML6) was fabricated in the binary system Ni-Ru. Figure 5a shows the compositional distribution and corresponding electrochemical activity of ML6. The binary composition spread exhibits a gradient with high Ru content in the top right and high Ni content in the bottom left of the library, resulting in an activity distribution symmetric along the diagonal. As expected, higher Ru contents correlate with increased activity. Figure 5b illustrates the updated composition space, where the newly acquired data points are positioned at one of the edges of the tetrahedron. The activities measured in ML6 surpass those of all previously tested compositions, and the observed trend indicates a global activity maximum at pure Ru. The binary library extends the total experimental coverage of the composition space by 0.8% to 19.4%.

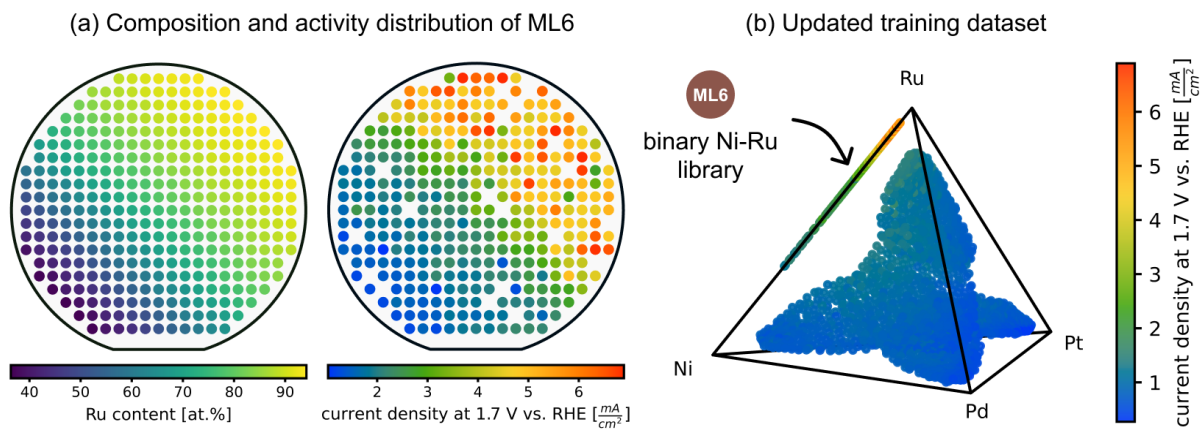

**Figure 5:** (a) Compositional distribution and electrochemical activity of the Ni-Ru materials library (ML6) and (b) the updated activity map for the quaternary composition space showing the new data points at the edge of the tetrahedron. The dataset covers 19.4% of the composition space. ML6 outperforms all previously measured libraries, with a global activity maximum suggested towards pure Ru.

In order to verify the activity maximum of pure Ru as the result of the search in the quaternary space, the activities of all four pure elements were determined. For each element, a thin film was prepared in the same sputter chamber using the same process temperature and pressure as for the materials libraries. Three individual SDC measurements were performed and the resulting LSVs were averaged. Figure 6 shows the LSVs and the updated activity dataset. Among the four elements, pure Ru exhibits the highest activity characterized by a rapid increase in current density, peaking just below 1.7 V vs. RHE. Ni shows moderate activity, with a less steep yet noticeable rise in current density at higher potentials, while Pd and Pt exhibit the lowest OER activity. The plateau observed in the LSV of Ru suggests that mass transport limitations or gas bubble formation may affect performance at elevated potentials. Incorporating the activities of the pure elements into the dataset, as illustrated in Figure 6b in the corners of the tetrahedron, reveals that the observations align well with the overall activity trend, with pure Ru being the best-performing catalyst in the Ni-Pd-Pt-Ru system.

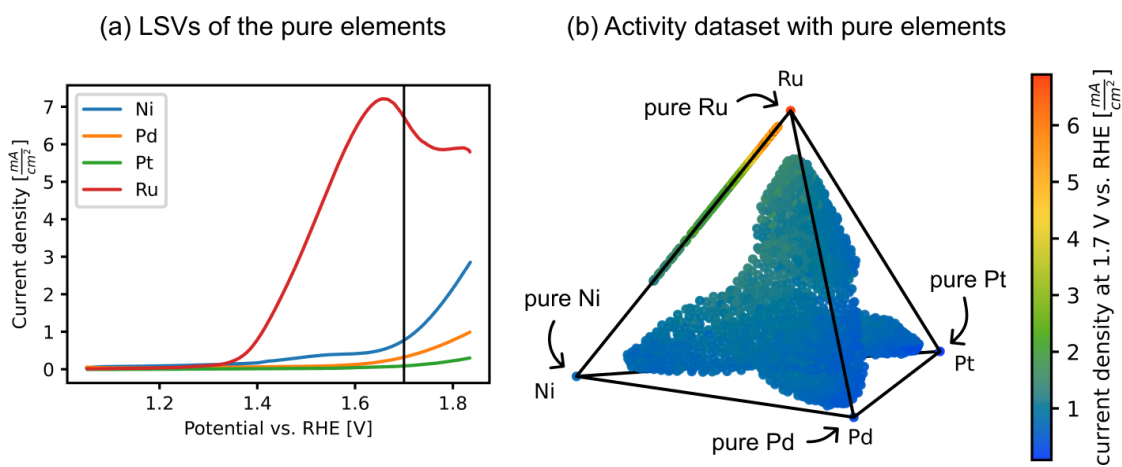

**Figure 6:** Measured LSVs of the pure elements (a) and updated activity distribution in the composition space (b). The vertical line in (a) highlights the potential of 1.7 V vs. RHE, which is plotted in (b). Pure Ru is the best catalyst in the system.

To further validate the global activity maximum of pure Ru in the Ni-Pd-Pt-Ru space, six additional materials libraries (ML7-12) were fabricated in regions with large distance from the already acquired compositions. Figure 7 shows the fabricated compositions alongside the activity distribution in the composition space. In total, more than 4000 compositions were acquired from the 12 materials libraries, covering 36.7% of the quaternary composition space. Five libraries were fabricated in all four ternary subspaces (ML8-12), while ML7 covers the compositional area around the highest uncertainty of the Gaussian process after the last training iteration. The additional libraries confirm the already found global activity maximum of pure Ru.

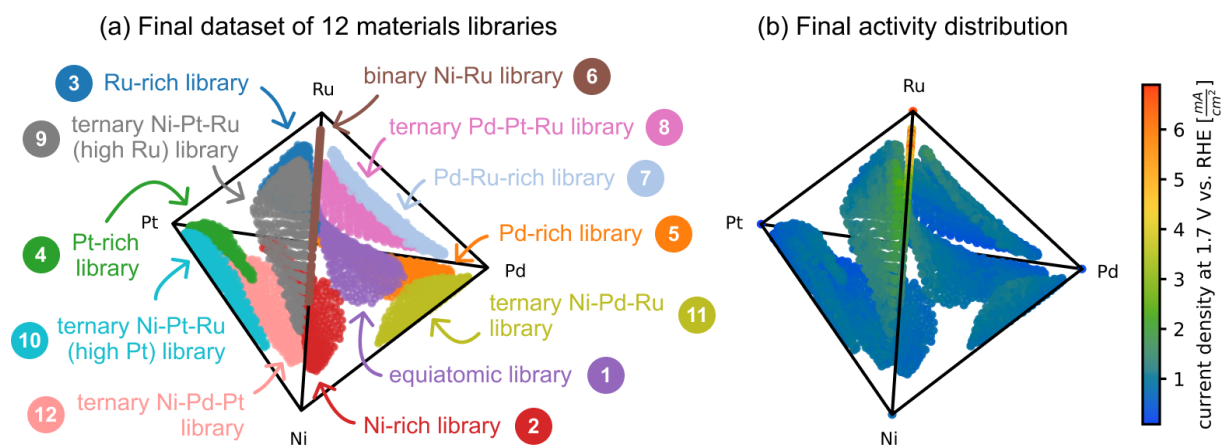

**Figure 7:** (a) Final dataset of 12 materials libraries covering 36.7% of the composition space. More than 4000 compositions were characterized. (b) Activity distribution in the composition space. Compared to other shown composition space plots, these figures are rotated to make all libraries visible.

**Discussion**

By combining combinatorial synthesis and high-throughput characterization with Bayesian optimization, the global activity maximum in the Ni-Pd-Pt-Ru composition space was successfully identified after only six materials libraries. Remarkably, exploration of only 20% of the composition space is needed to identify the optimal composition. This demonstrates that despite the vast size of multinary composition spaces, efficient exploration can significantly reduce experimental effort. Given that the fabrication and characterization of a single materials library requires approximately two days, the total time investment for this exploration can be estimated to 12 days. The six additional libraries allowed for a systematic validation, confirming pure Ru as the global activity maximum in the system. This finding aligns with Ru's well-established role as one of the most efficient OER catalysts. $RuO_2$, which forms under reaction conditions, is located close to the top of the volcano plot [18] and therefore features nearly optimal binding energies for the reaction intermediates. The incorporation of Ni, Pd and Pt to form compositionally complex solid solutions could have led to a better-performing catalyst. However, in the specific system which we investigated here, compositional complexity did not enhance catalytic activity compared to the best-performing constituent element Ru. Instead, the additional elements likely disrupted the nearly optimal binding energy characteristic for Ru, leading to the observed activity trend in the quaternary system Ni-Pd-Pt-Ru. Additionally, the statistical replacement of active Ru centers through the incorporation of less active elements could have also contributed to the reduction of the overall activity.

Although the global activity maximum was identified with relatively small experimental effort, the twelve materials libraries are likely insufficient to accurately predict the remaining composition space. This limitation becomes evident when performing a train-test split, where the model is trained on eleven libraries and tasked with predicting the excluded one. Figure 8a shows the prediction error of each excluded materials library. While the model captures the activity trend for each library, the absolute predicted values deviate significantly, resulting in a relatively high mean absolute error. This is particularly notable as the median activity range of the libraries is 1.2 mA/cm$^2$. The Ni-Ru library (ML6) exhibits the highest prediction error by a wide margin. A comparison of Figure 8c and Figure 5a reveals that while the model correctly predicts the trend of increasing activity with higher Ru content for this library, lower activity values are consistently overestimated, whereas higher activities are significantly underestimated.

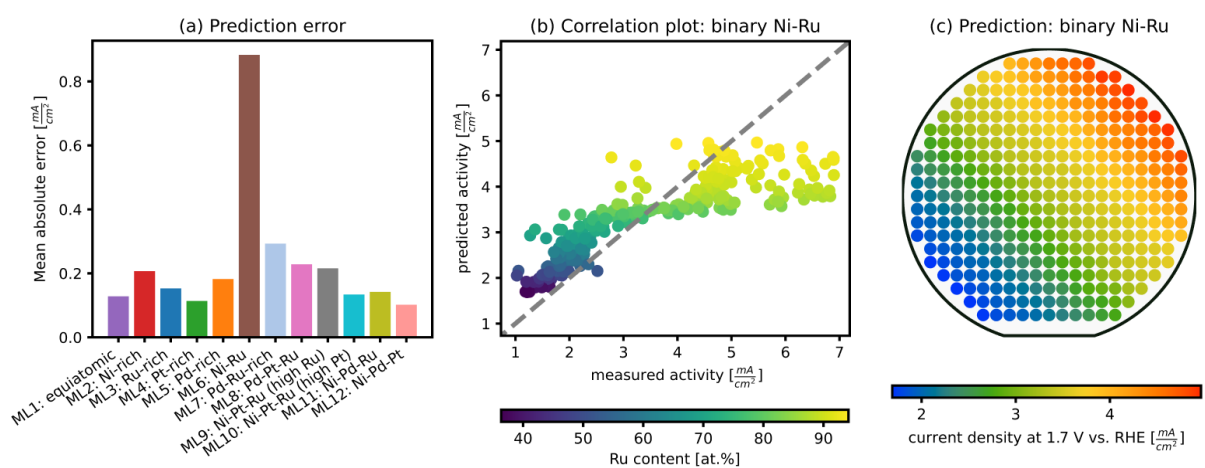

**Figure 8:** Evaluation of the model's performance for predicting individual excluded materials libraries in a train-test split. Mean absolute error of the predictions for each excluded library (a) and correlation plot (b) as well as the predicted activity (c) for the Ni-Ru library, which shows the highest prediction error.

Several factors contribute to this discrepancy. One major limitation could be missing features in the training data to successfully predict the catalytic activity. As the activity is predominantly a surface-related property, compositions acquired by EDX only serve as a proxy for the surface composition in the predictions. However, the surface composition can differ from the volume of the film due to surface segregation. While X-ray photoelectron spectroscopy (XPS) could provide insights into the surface composition, its measurement and analysis time remains significantly longer than other high-throughput techniques like EDX, making it impractical for measuring entire materials libraries. Additionally, surface morphology and crystal structure also influence catalytic performance. The XRD analysis reveals that most of the composition space shows an fcc crystal structure. However, in the Ru-rich regions of ML7 and ML9, particularly in the Ru-rich library (ML3) and the binary Ni-Ru library (ML6), the crystal structure transitions into hexagonal closed packed (hcp) as the dominant phase. A figure highlighting this transition can be found in the supporting information (see Figure S-2). This change in crystal structure can explain why the Ni-Ru library in particular could not be predicted accurately, as the model is tasked with learning the activity distribution while the crystal structure remains unaccounted for, thereby reducing its predictive accuracy. Ensuring that the experiments are conducted in single-phase compositional regions could help to reduce the complexity and reduce the influence of the crystal structure on model performance in future studies. Similarly, incorporating high-throughput electrochemical surface area (ECSA) measurements [19] could address the influence of morphology by weighting the activity appropriately.

Apart from missing features, the twelve libraries inside the training dataset collectively cover only about 30% of the quaternary composition space, providing insufficient data for accurate predictions. The sparse data coverage limits the model's ability to generalize, particularly in unexplored regions. Libraries farther away from the majority of observations, such as the binary Ni-Ru (ML6) or the Pd-Ru-rich library (ML7) show a much greater prediction error when excluded from the training data compared to libraries in more densely sampled compositional regions, such as the equiatomic (ML1) or Pt-rich libraries (ML4). As a non-parametric algorithm, a Gaussian process only makes very few assumptions about the underlying function, e.g. via the chosen kernel function. While this flexibility makes it versatile, it also weakens its predictive performance in regions with sparse or no training data. This was addressed in this study by initializing the Bayesian optimization with five materials libraries covering a broader compositional space instead of a single one, making the initial predictions more robust.

While the model struggles with accurately predicting entire materials libraries left out of the training dataset, its performance remains relatively consistent when data points are removed from each library. By excluding every second, third, and higher fraction of data points, the model still achieves a coefficient of determination ($R^2$) exceeding 90%, even when trained on $1/20^{th}$ of the available data. This is visualized in Figure 9, which shows the $R^2$ values as a function of the fraction of data points used for training. When the model is trained on the full dataset, it achieves an $R^2$ of 95%, demonstrating its ability to generalize the training data.

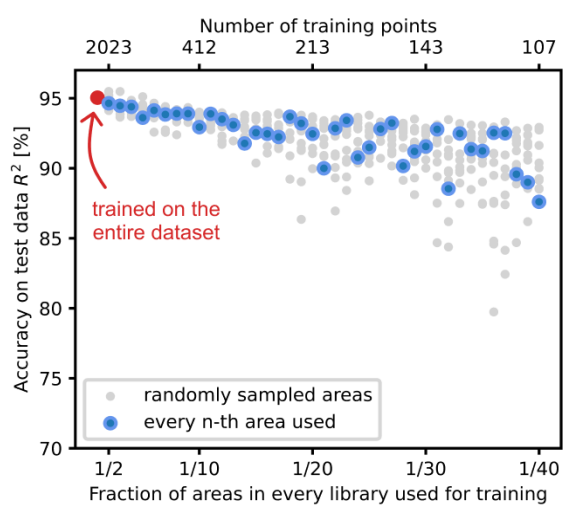

**Figure 9:** Effect of reducing the fraction of training data on the model's predictive accuracy (quantified by the coefficient of determination $R^2$). Blue data points denote the accuracy when trained on every $n^{th}$ measurement area of each library, while grey points indicate the accuracy when trained on randomly sampled areas. The model achieves an $R^2$ of 95% when trained on the full dataset, demonstrating its ability to generalize the training data. Even when trained on as little as $1/20^{th}$ of the available data, the model maintains an $R^2$ exceeding 90%, indicating a modest decrease in accuracy. The equivalent dataset size is shown on the secondary axis. The size depends on the number of areas excluded manually due to SDC measurement errors.

When fractions of the dataset are used for training, the accuracy decreases modestly as the fraction becomes smaller. For smaller fractions, the variability of $R^2$ increases, as the selection of specific data points becomes more critical with reduced dataset size. Consequently, the reliability of the shown relationship decreases for smaller fractions. This is highlighted by grey data points, showing the accuracy when the selection of measurement areas was performed randomly. However, the figure suggests that measuring up to one-fifth of the 342 pre-defined areas on each library results is a tolerable decrease in accuracy while significantly improving experimental efficiency. The slight decrease in accuracy when leaving out data points can be attributed to the fact that the compositions and activities within each library vary only by small magnitudes relative to the size of the overall composition space and the distances between different libraries. This suggests that the fabrication of more libraries should be favored over the characterization of all measurement areas on a library. This is expected to become even more significant when exploring quinary composition spaces, as the local compositional gradients within a single library are even smaller relative to the overall composition space.

**Conclusion**

Bayesian optimization can guide researchers efficiently through high-dimensional search spaces. In case of the explored Ni-Pd-Pt-Ru system, the approach enabled the efficient identification of the global activity maximum after the fabrication and characterization of only six materials libraries, covering approximately 20% of the quaternary composition space. Pure Ru was confirmed to be the most active OER catalyst in this system. Six additional libraries were fabricated, and their characterization validated the activity trend.

Despite the experimental effort of investigating 4000 compositions, the analysis showed that achieving comprehensive coverage of a quaternary composition space is experimentally costly, even when using combinatorial synthesis techniques and high-throughput characterization. The explored regions correspond to 37% of the quaternary composition space, highlighting the challenge of achieving complete coverage. Due to this sparse data coverage, the employed Gaussian process model showed limited performance when extrapolating to unexplored regions. Libraries located farther away from the training data exhibited higher prediction errors. Nonetheless, the model showed high robustness when interpolating within observed regions of the composition space. Iteratively removing points from the training data showed that even when trained on $1/20^{th}$ of the available data, the model maintained a high predictive accuracy of more than 90%. This suggests that with the advantage of co-sputtering compositional gradients the exploration efficiency can be significantly improved compared to one-at-a-time experiments. However, since not the entire compositional gradient is needed for accurate predictions, the number of measurements on a materials library should be reduced for future composition space explorations. In turn, the fabrication of more libraries should be preferred in order to sample more regions of the composition space.

This work demonstrates the potential of combining combinatorial synthesis with Bayesian optimization for the efficient exploration of multinary composition spaces and the reduction of experimental costs, providing a foundation for extending this methodology to even more complex materials systems. This is important for the goal of understanding and finding optimal electrocatalysts for industrial applications, in the hydrogen economy and beyond.

**Supporting information**

The authors have cited additional references within the supporting information [20–24].

**Data availability**

The dataset of this publication is accessible at Zenodo under https://doi.org/10.5281/zenodo.14891703

**Author contributions**

F. Thelen: conceptualization, sample fabrication, measurement, data curation, formal analysis, investigation, software, validation, visualization, supervision, writing – original draft (lead), writing – review & editing. R. Zehl: conceptualization, sample fabrication, measurement, investigation, supervision, writing – original draft (support), writing – review & editing. R. Zerdoumi: measurement, writing – original draft. J. L. Bürgel: sample fabrication, measurement, writing – review & editing. L. Banko: conceptualization, writing – review & editing. W. Schuhmann: supervision, resources, writing – review & editing. A. Ludwig: supervision, resources, writing – review & editing, project administration.

**Conflict of interest**

There are no conflicts to declare.

**Acknowledgements**

This work was partially supported from different projects. A. Ludwig, F. Thelen and J. L. Bürgel acknowledge funding from the European Union through the European Research Council (ERC) Synergy Grant project 101118768 (DEMI). Funding by the Deutsche Forschungsgemeinschaft (DFG, German Research Foundation) CRC 1625, project number 506711657, subproject A02 is acknowledged by R. Zehl and A. Ludwig, as well as subproject C01 by W. Schuhmann. R. Zerdoumi and W. Schuhmann acknowledge financial support by the ERC under the European Union's Horizon 2020 research and innovation programme (CasCat [833408]). The authors acknowledge using ZGH infrastructure for measurements (Scanning Electron Microscope and X-Ray Diffractometer).

**References**

[1] T. Löffler, A. Ludwig, J. Rossmeisl, W. Schuhmann, What Makes High-Entropy Alloys Exceptional Electrocatalysts?, Angewandte Chemie International Edition 60 (2021) 26894–26903. https://doi.org/10.1002/anie.202109212.

[2] M.L. Green, C.L. Choi, J.R. Hattrick-Simpers, A.M. Joshi, I. Takeuchi, S.C. Barron, E. Campo, T. Chiang, S. Empedocles, J.M. Gregoire, A.G. Kusne, J. Martin, A. Mehta, K. Persson, Z. Trautt, J. Van Duren, A. Zakutayev, Fulfilling the promise of the materials genome initiative with high-throughput experimental methodologies, Appl Phys Rev 4 (2017) 011105. https://doi.org/10.1063/1.4977487.

[3] J.T. Gudmundsson, D. Lundin, Introduction to magnetron sputtering, in: D. Lundin, T. Minea, J.T. Gudmundsson (Eds.), High Power Impulse Magnetron Sputtering, Elsevier, 2020: pp. 1–48. https://doi.org/https://doi.org/10.1016/B978-0-12-812454-3.00006-1.

[4] A. Ludwig, Discovery of new materials using combinatorial synthesis and high-throughput characterization of thin-film materials libraries combined with computational methods, NPJ Comput Mater 5 (2019) 70. https://doi.org/10.1038/s41524-019-0205-0.

[5] K. Sliozberg, D. Schäfer, T. Erichsen, R. Meyer, C. Khare, A. Ludwig, W. Schuhmann, High-Throughput Screening of Thin-Film Semiconductor Material Libraries I: System Development and Case Study for Ti-W-O, ChemSusChem 8 (2015) 1270–1278. https://doi.org/10.1002/cssc.201402917.

[6]     L. Banko, O.A. Krysiak, J.K. Pedersen, B. Xiao, A. Savan, T. Löffler, S. Baha, J. Rossmeisl, W. Schuhmann, A. Ludwig, Unravelling Composition–Activity–Stability Trends in High Entropy Alloy Electrocatalysts by Using a Data-Guided Combinatorial Synthesis Strategy and Computational Modeling, Adv Energy Mater 12 (2022). https://doi.org/10.1002/aenm.202103312.

[7]     B. Settles, Active Learning Literature Survey, Computer Sciences Technical Report 1648 (2014).

[8]     E. Brochu, V.M. Cora, N. De Freitas, A Tutorial on Bayesian Optimization of Expensive Cost Functions, with Application to Active User Modeling and Hierarchical Reinforcement Learning, ArXiv Preprint (2010). https://doi.org/10.48550/arXiv.1012.2599.

[9]     P.I. Frazier, A Tutorial on Bayesian Optimization, ArXiv Preprint (2018). https://doi.org/10.48550/arXiv.1807.02811.

[10]    X. Yue, Y. Wen, J.H. Hunt, J. Shi, Active Learning for Gaussian Process Considering Uncertainties With Application to Shape Control of Composite Fuselage, IEEE Transactions on Automation Science and Engineering 18 (2021) 36–46. https://doi.org/10.1109/TASE.2020.2990401.

[11]    A.G. Kusne, H. Yu, C. Wu, H. Zhang, J. Hattrick-Simpers, B. DeCost, S. Sarker, C. Oses, C. Toher, S. Curtarolo, A. V. Davydov, R. Agarwal, L.A. Bendersky, M. Li, A. Mehta, I. Takeuchi, On-the-fly closed-loop materials discovery via Bayesian active learning, Nat Commun 11 (2020) 5966. https://doi.org/10.1038/s41467-020-19597-w.

[12]    J.K. Pedersen, C.M. Clausen, O.A. Krysiak, B. Xiao, T.A.A. Batchelor, T. Löffler, V.A. Mints, L. Banko, M. Arenz, A. Savan, W. Schuhmann, A. Ludwig, J. Rossmeisl, Bayesian Optimization of High-Entropy Alloy Compositions for Electrocatalytic Oxygen Reduction**, Angewandte Chemie 133 (2021) 24346–24354. https://doi.org/10.1002/ange.202108116.

[13]    T. Lookman, P. V Balachandran, D. Xue, R. Yuan, Active learning in materials science with emphasis on adaptive sampling using uncertainties for targeted design, NPJ Comput Mater 5 (2019) 21. https://doi.org/10.1038/s41524-019-0153-8.

[14]    V.A. Mints, J.K. Pedersen, A. Bagger, J. Quinson, A.S. Anker, K.M.Ø. Jensen, J. Rossmeisl, M. Arenz, Exploring the Composition Space of High-Entropy Alloy Nanoparticles for the Electrocatalytic $H_2$/CO Oxidation with Bayesian Optimization, ACS Catal 12 (2022) 11263–11271. https://doi.org/10.1021/acscatal.2c02563.

[15]    B. Burger, P.M. Maffettone, V. V. Gusev, C.M. Aitchison, Y. Bai, X. Wang, X. Li, B.M. Alston, B. Li, R. Clowes, N. Rankin, B. Harris, R.S. Sprick, A.I. Cooper, A mobile robotic chemist, Nature 583 (2020) 237–241. https://doi.org/10.1038/s41586-020-2442-2.

[16]    B.P. MacLeod, F.G.L. Parlane, T.D. Morrissey, F. Häse, L.M. Roch, K.E. Dettelbach, R. Moreira, L.P.E. Yunker, M.B. Rooney, J.R. Deeth, V. Lai, G.J. Ng, H. Situ, R.H. Zhang, M.S. Elliott, T.H. Haley, D.J. Dvorak, A. Aspuru-Guzik, J.E. Hein, C.P. Berlinguette, Self-driving laboratory for accelerated discovery of thin-film materials, Sci Adv 6 (2020). https://doi.org/10.1126/sciadv.aaz8867.

[17] C.M. Clausen, O.A. Krysiak, L. Banko, J.K. Pedersen, W. Schuhmann, A. Ludwig, J. Rossmeisl, A Flexible Theory for Catalysis: Learning Alkaline Oxygen Reduction on Complex Solid Solutions within the Ag−Pd−Pt−Ru Composition Space**, Angewandte Chemie International Edition 62 (2023). https://doi.org/10.1002/anie.202307187.

[18] S. Trasatti, Electrocatalysis by oxides — Attempt at a unifying approach, J Electroanal Chem Interfacial Electrochem 111 (1980) 125–131. https://doi.org/10.1016/S0022-0728(80)80084-2.

[19] E. Suhr, O.A. Krysiak, V. Strotkötter, F. Thelen, W. Schuhmann, A. Ludwig, High-Throughput Exploration of Structural and Electrochemical Properties of the High-Entropy Nitride System (Ti–Co–Mo–Ta–W)N, Adv Eng Mater 25 (2023). https://doi.org/10.1002/adem.202300550.

[20] F. Thelen, R. Zehl, J.L. Bürgel, D. Depla, A. Ludwig, A Python-Based Approach to Sputter Deposition Simulations in Combinatorial Materials Science, (2024).

[21] A.G. de G. Matthews, M. van der Wilk, T. Nickson, K. Fujii, A. Boukouvalas, P. Leon-Villagra, Z. Ghahramani, J. Hensman, GPflow: A Gaussian process library using TensorFlow, Journal of Machine Learning Research 18 (2017) 1–6. http://jmlr.org/papers/v18/16-537.html (accessed May 24, 2023).

[22] C.E. Rasmussen, C.K.I. Williams, Gaussian Processes for Machine Learning, The MIT Press, Massachusetts, 2006.

[23] F. Thelen, L. Banko, R. Zehl, S. Baha, A. Ludwig, Speeding up high-throughput characterization of materials libraries by active learning: autonomous electrical resistance measurements, Digital Discovery 2 (2023) 1612–1619. https://doi.org/10.1039/D3DD00125C.

[24] D. Duvenaud, The Kernel Cookbook: Advice on Covariance functions, (2014). https://www.cs.toronto.edu/~duvenaud/cookbook/ (accessed May 5, 2022).

# Supporting Information

**Experimental section**

Twelve materials libraries in the system Ni-Pd-Pt-Ru were fabricated at room temperature in a four-cathode co-sputter system (AJA International Polaris). The depositions were performed on 10 cm diameter single-side polished sapphire wafers (SITUS Technicals, c-plane orientation). Three DC power supplies (2x DCXS-750, 1x DCXS-1500) and a 0313 GTC RF unit (all AJA International) were used in power-control mode. The gas flow rate was 80 sccm Ar at a pressure of 0.5 Pa. An overview of the sputter parameters can be found in Table S-1. The deposition powers were determined based on preliminary sputter rate determinations and Monte Carlo sputter deposition simulations using pySIMTRA [1]. Due to the high sputter yield of Pd, the cathode equipped with Pd was assigned to an RF power supply. The single element thin films were deposited under identical conditions with a power of 35 W.

**Table S-1:** Power setpoint values used for each deposition and cathode.

|         |                    | Ni | Pd | Pt | Ru |
|---------|--------------------|----|----|----|----|
|         |                    | Power [W] | | | |
| Library |                    | DC | RF | DC | DC |
| ML1     | Equiatomic         | 32 | 32 | 15 | 30 |
| ML2     | Ni-rich            | 45 | 11 | 6  | 8  |
| ML3     | Ru-rich            | 5  | 7  | 5  | 35 |
| ML4     | Pt-rich            | 9  | 7  | 35 | 5  |
| ML5     | Pd-rich            | 5  | 60 | 5  | 5  |
| ML6     | Ni-Ru              | 15 | -  | -  | 40 |
| ML7     | Pd-Ru-rich         | 9  | 80 | 5  | 80 |
| ML8     | Pd-Pt-Ru           | -  | 40 | 19 | 34 |
| ML9     | Ni-Pt-Ru (high Ru) | 42 | -  | 8  | 60 |
| ML10    | Ni-Pt-Ru (high Pt) | 45 | -  | 55 | 7  |
| ML11    | Ni-Pd-Ru           | 60 | 100| -  | 7  |
| ML12    | Ni-Pd-Pt           | 52 | 14 | 30 | -  |

The chemical composition was determined by automatically measuring 342 areas on each library using energy-dispersive X-ray spectroscopy (EDX) in a scanning electron microscope (SEM, JEOL L 7200F) equipped with an EDS detector (Oxford AZtecEnergy X-MaxN 80 mm$^2$). The electrochemical activity was measured with an automated scanning droplet cell (SDC) setup in a conventional three-electrode configuration, capable of automatically approaching all 342 areas as well. A Pt wire served as a counter electrode and a Ag|AgCl|3M KCl electrode as a reference electrode. The electrochemical cell is formed upon pressing the SDC tip onto the library, creating a circular contact area of 0.00735 cm$^2$, defined by the 1 mm tip opening. The applied force is monitored using a force sensor integrated into the tip holder. The OER activity was measured using linear sweep voltammetry in the potential range of 1.0 to 1.8 V vs. RHE at a scan rate of 10 mV s$^{-1}$. The measured current was normalized to the geometric surface area of the tip and no iR drop compensation was performed. All potentials were calculated and reported versus RHE using the following equation, where $E_{(Ag|AgCl|3M\ KCl)}$ is the measured potential versus Ag|AgCl|3M KCl:

$$E_{RHE} = E_{(Ag|AgCl|3M\ KCl)} + 0.210 + 0.059\ pH$$

In order to obtain a one-dimensional activity measure from the LSVs the current density at a potential of 1.7 V (vs. RHE) was extracted.

For guiding the materials library synthesis experiments through the Ni-Pd-Pt-Ru composition space, a Gaussian process implemented in GPflow [2] was used in a Bayesian optimization loop in order to find the composition with the highest catalytic activity for the OER. The Gaussian process was supplied with the measured compositions as input and the activity measure as output training data. Since the shape of the multi-dimensional function, which should be learned by the Gaussian process, is unknown in advance, the Matérn52 kernel was selected. It can provide a more flexible fit due to a higher number of hyperparameters compared to the most commonly used squared exponential kernel [3]. Also, it was found to be more performant on a similar study on co-sputtered materials libraries [4]. The predictions were done on the entire composition space sampled at 5 at.% steps. The compositions to fabricate next were determined by maximizing the expected improvement acquisition function [5] based on the predicted activity and the uncertainty of the Gaussian process.

**Coverage determination**

To quantify the exploration progress in covering the composition space of interest, a coverage metric was defined, which specifies the amount of an $n$-dimensional composition space covered by a set of $p$-compositions $C_f = \{c_{f,i}\}_{i=1}^{p}$. A second set of compositions $C_s = \{c_{s,i}\}_{i=1}^{q}$ is obtained by sampling $q$-compositions from the composition space with a step size of $t\ at.\%$ according to

$$q = \binom{T-n-1}{n-1}$$

with $T = 100 \cdot t^{-1}$ being the number of partitions along each composition space axis. This is shown in Figure S-1 for a ternary materials library, which covers 30.3% of the ternary composition space.

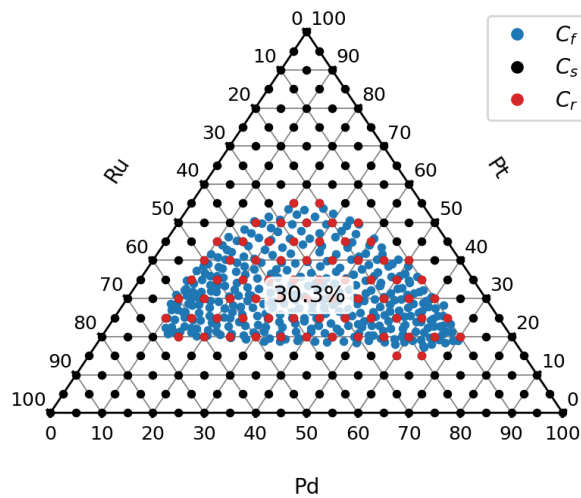

**Figure S-1:** Plot visualizing the coverage calculation. The compositions of a materials library obtained by EDX are shown in blue. All possible compositions in 5 at. % steps are shown in black and all nearest neighbors to the measured compositions in red. The coverage is defined as the ratio of the number of points in $C_r$ and the total number of combinations in $C_r$, which corresponds to 70/231.

For a quaternary composition space sampled with steps of $5\,at.\%$, this corresponds to $q = 1771$ compositions. When there is a set of $r$-nearest neighbors of $C_s$ in $C_f$, called $C_r$, the coverage $cov$ of the compositions $C_f$ is defined as the ratio of the number of nearest neighbors $r$ to the total number of sampled compositions $q$:

$$cov = \frac{r}{q} \cdot 100\ [\%]$$

**Analysis of the crystal structure**

To illustrate the change in crystal structure across the composition space, the hcp (101) peak was tracked in the two-theta range of 42° - 46°, as shown in Figure S-2b. This peak was observed exclusively in libraries ML3, ML6, ML7 and ML9, with the highest intensity detected in ML6. Figure S-2a shows exemplary diffractograms highlighting this trend. Towards higher Ru content, the hcp peaks become increasingly visible, indicating a structural transition. In contrast, throughout the remainder of the composition space, only an fcc phase was detected. The observed trend suggests that Ni-Ru combinations favor the formation of the hcp phase, whereas mixtures with other elements exhibit the hcp structure only at high Ru concentrations.

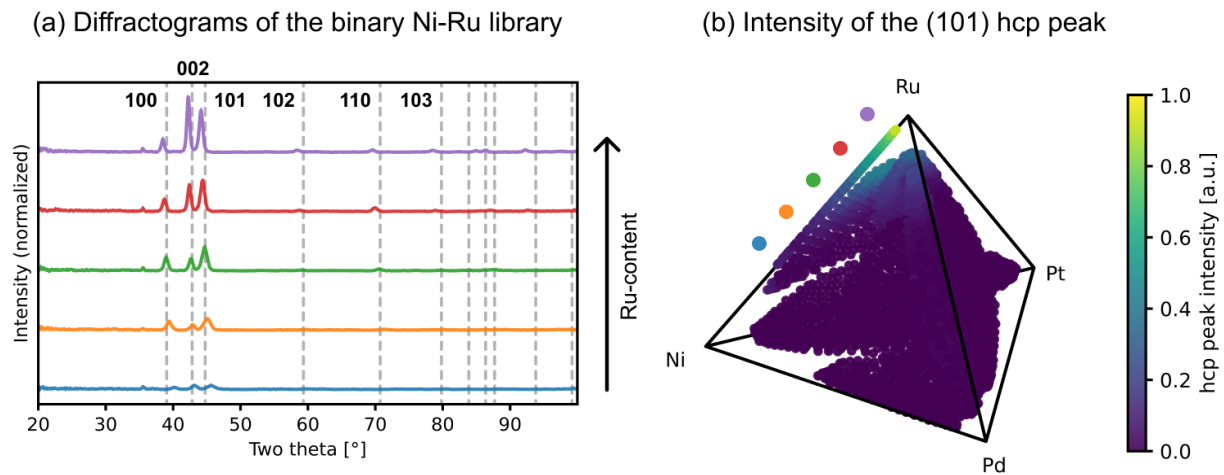

**Figure S-2:** (a) Exemplary XRD diffractograms of the binary Ni-Ru library along the compositional gradient. (b) Intensity map of the hcp (101) peak in the composition space. The peak is only visible in the libraries ML3, ML6, ML7 and ML9. The intensity of this peak increases with higher Ru content, with the strongest signal observed in ML6, while the rest of the composition space shows only an fcc phase.

**References**

[1]   F. Thelen, R. Zehl, J.L. Bürgel, D. Depla, A. Ludwig, A Python-Based Approach to Sputter Deposition Simulations in Combinatorial Materials Science, (2024).

[2]   A.G. de G. Matthews, M. van der Wilk, T. Nickson, K. Fujii, A. Boukouvalas, P. Leon-Villagra, Z. Ghahramani, J. Hensman, GPflow: A Gaussian process library using TensorFlow, Journal of Machine Learning Research 18 (2017) 1–6. http://jmlr.org/papers/v18/16-537.html (accessed May 24, 2023).

[3]   C.E. Rasmussen, C.K.I. Williams, Gaussian Processes for Machine Learning, The MIT Press, Massachusetts, 2006.

[4]   F. Thelen, L. Banko, R. Zehl, S. Baha, A. Ludwig, Speeding up high-throughput characterization of materials libraries by active learning: autonomous electrical resistance measurements, Digital Discovery 2 (2023) 1612–1619. https://doi.org/10.1039/D3DD00125C.

[5]   D. Duvenaud, The Kernel Cookbook: Advice on Covariance functions, (2014). https://www.cs.toronto.edu/~duvenaud/cookbook/ (accessed May 5, 2022).

# Dataset figures

**Figures S-3-14:** LSVs and activity gradients of all 12 materials libraries.

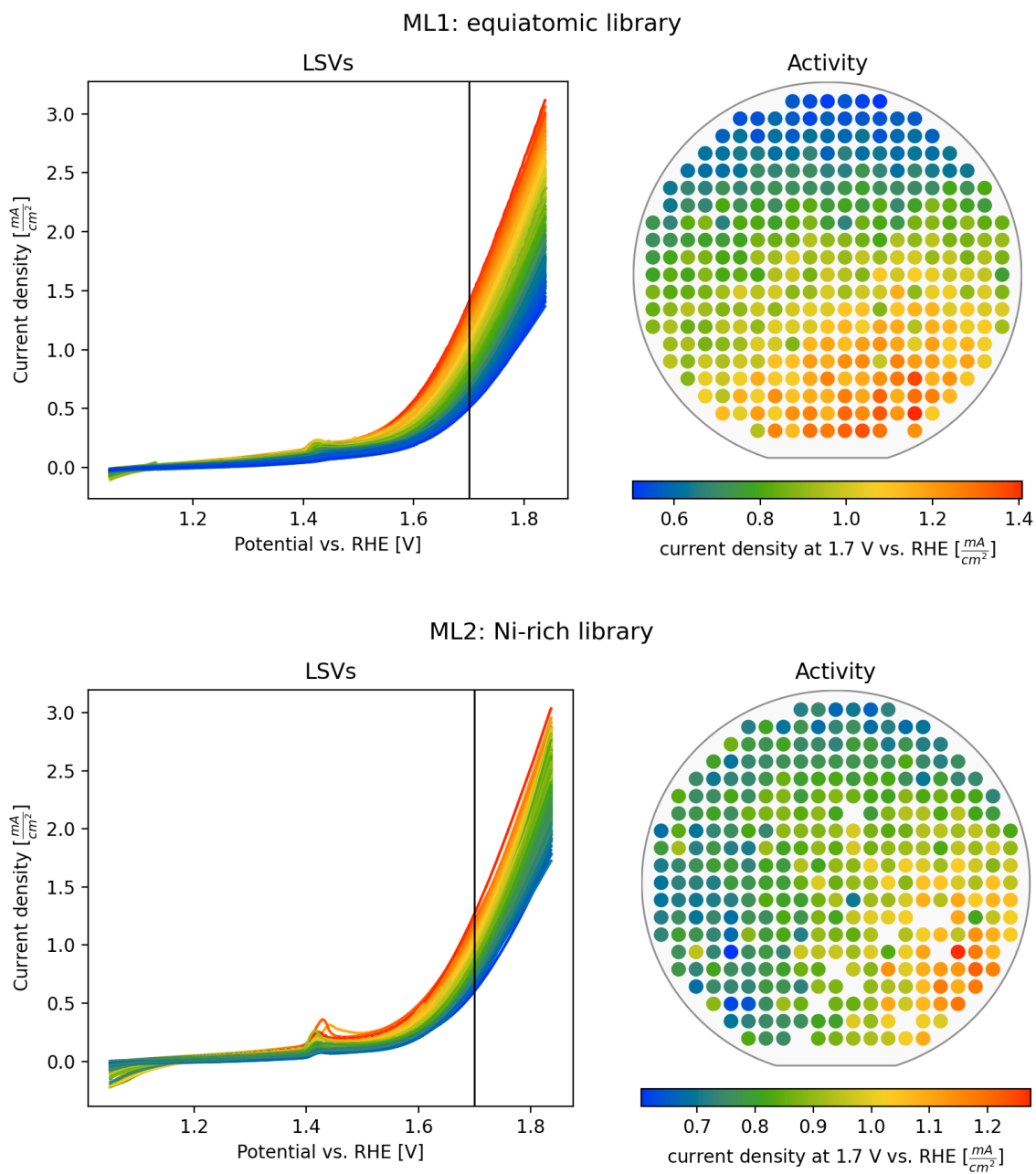

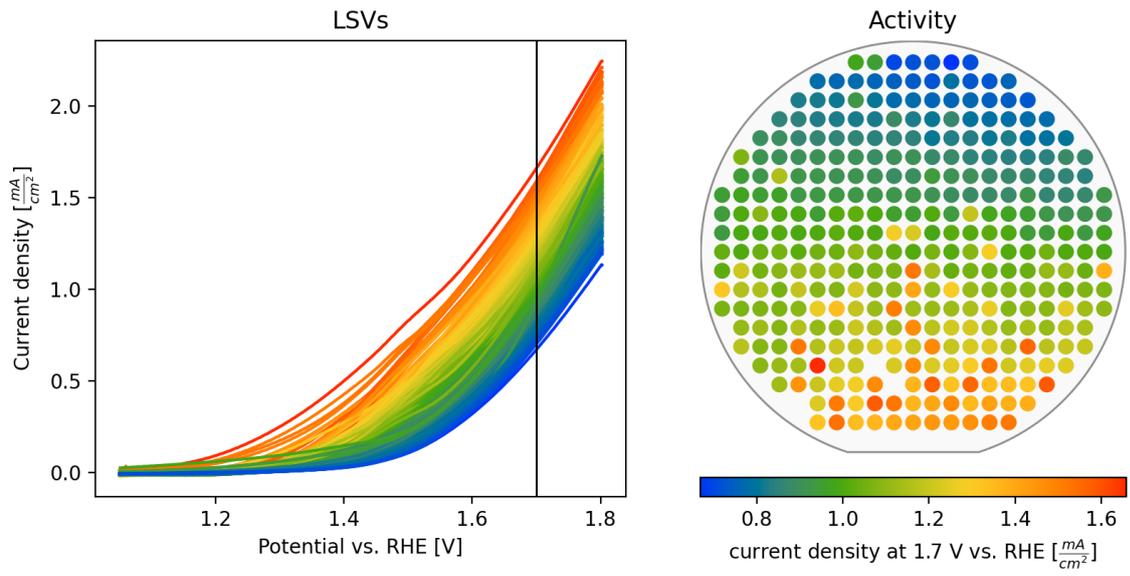

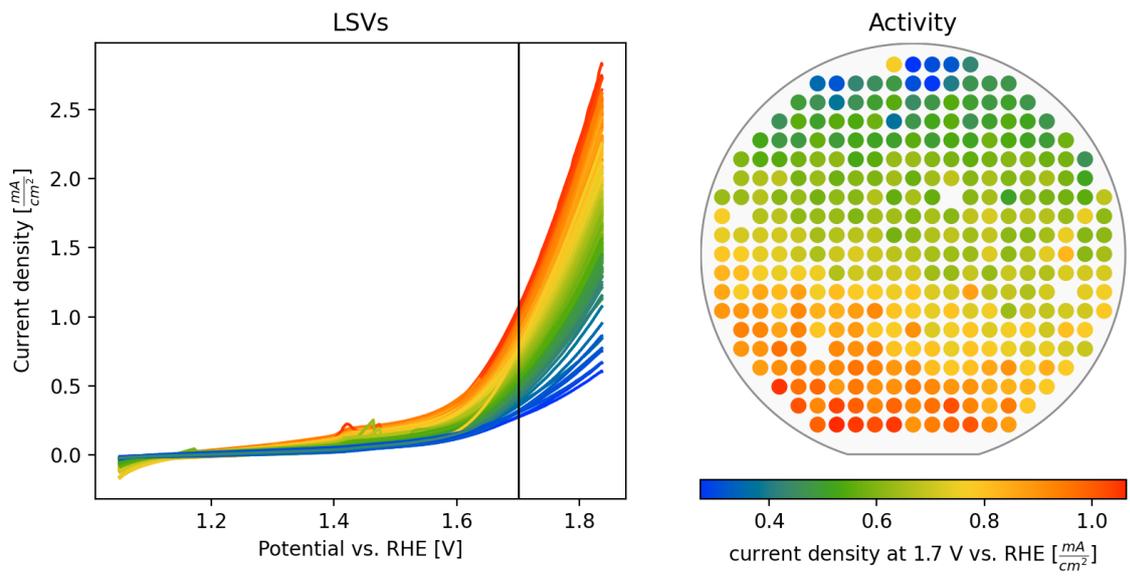

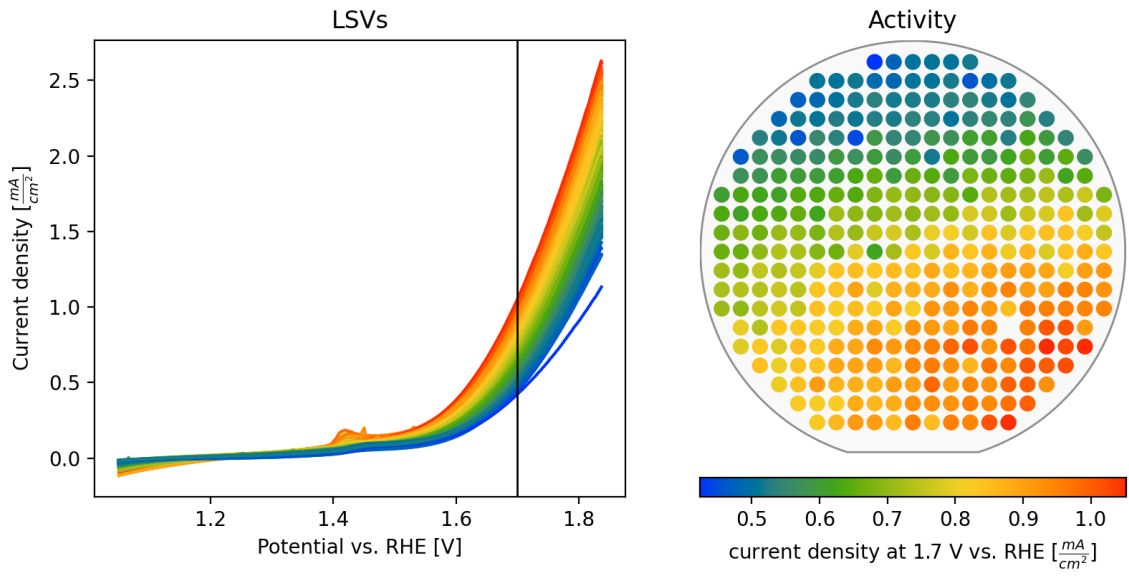

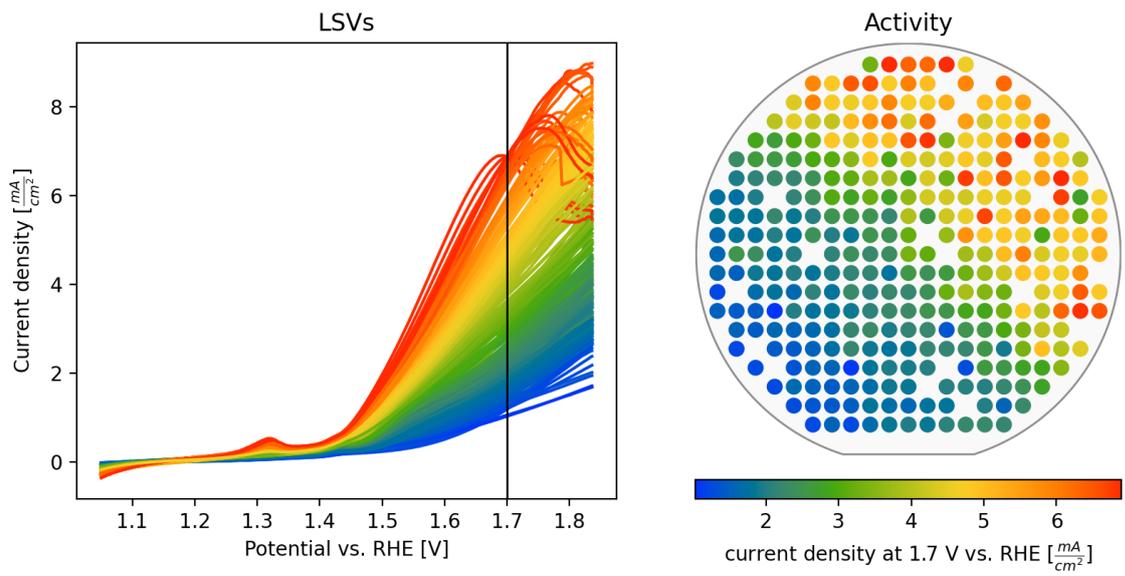

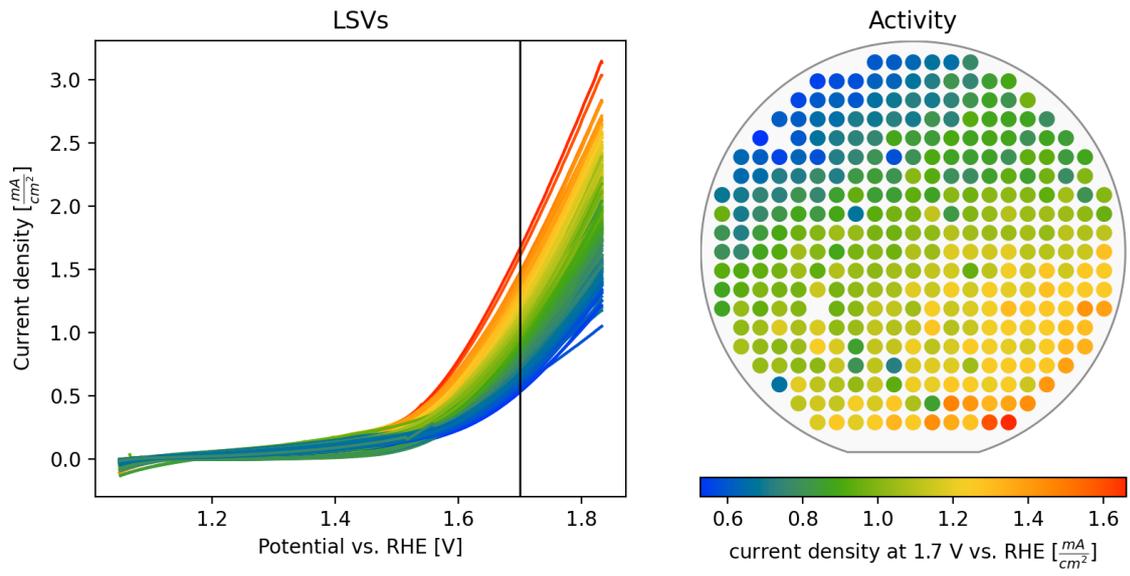

ML7: Pd-Ru-rich library

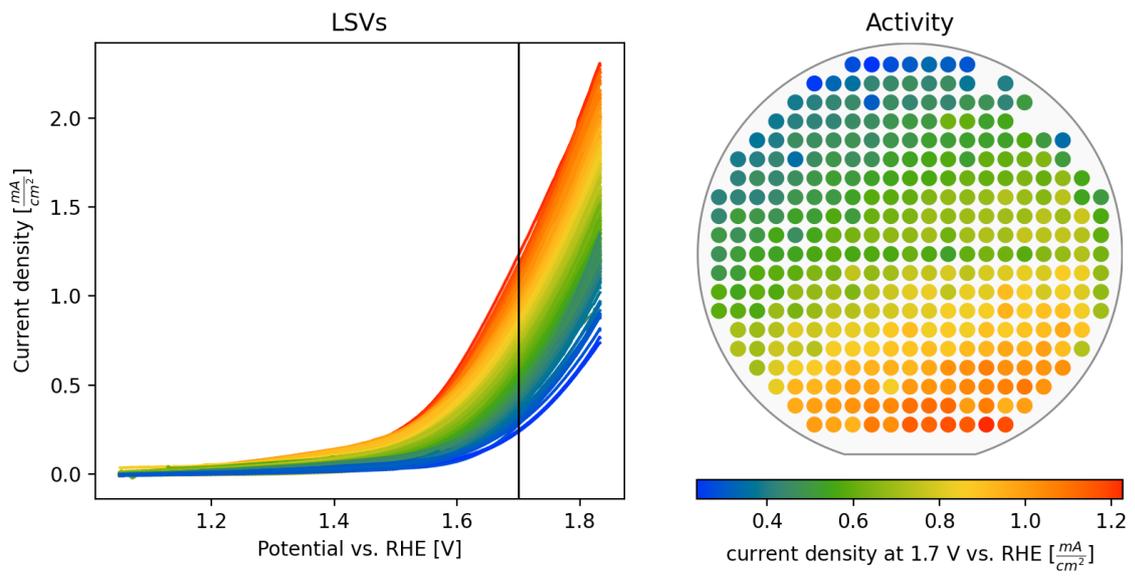

ML8: ternary Pd-Pt-Ru library

## ML9: ternary Ni-Pt-Ru (high Ru) library

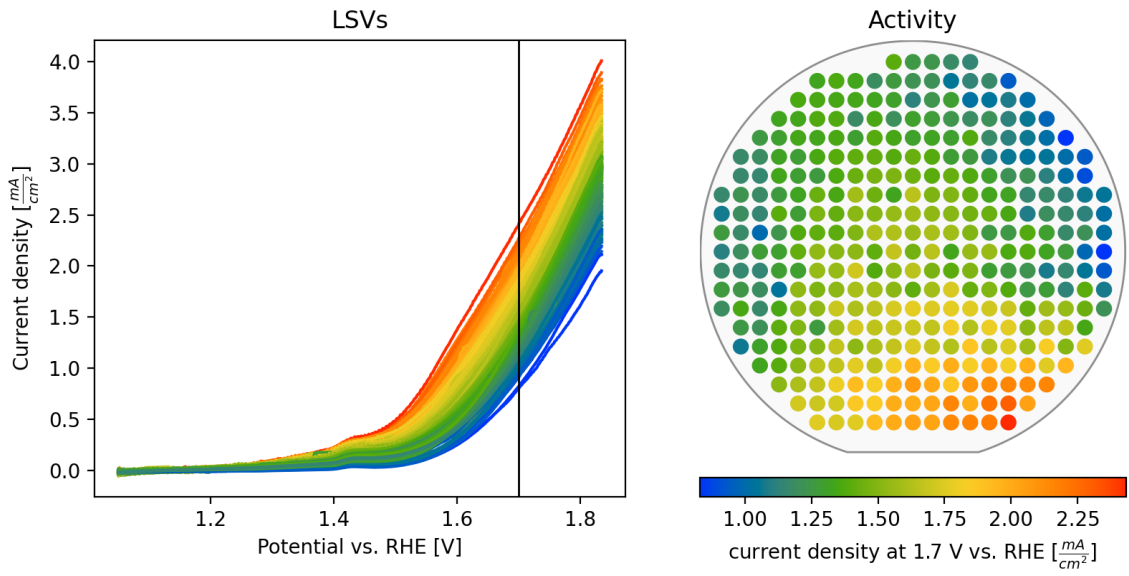

## ML10: ternary Ni-Pt-Ru (high Pt) library

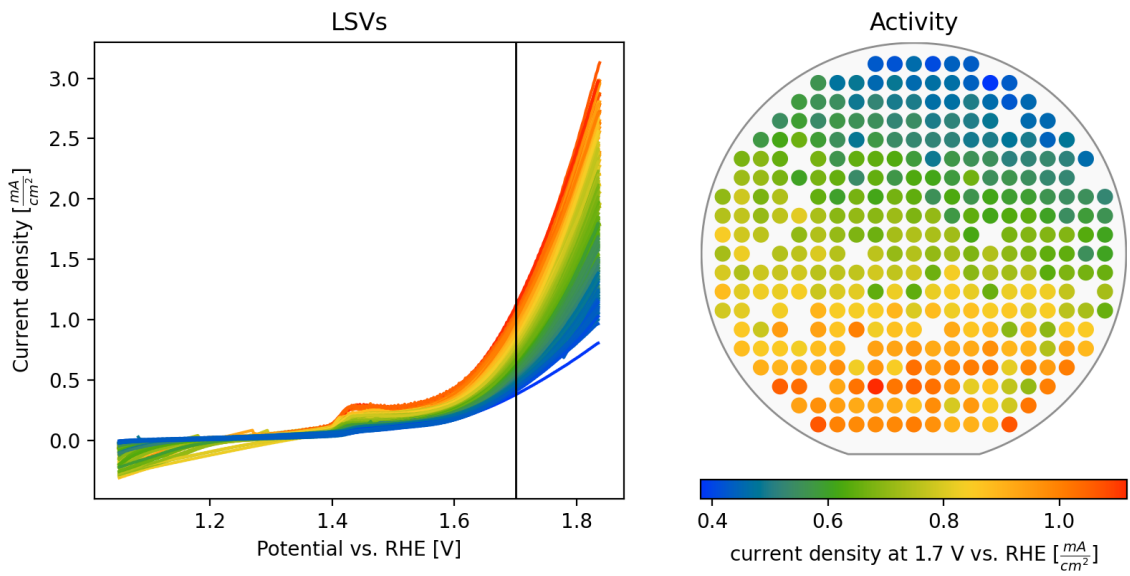

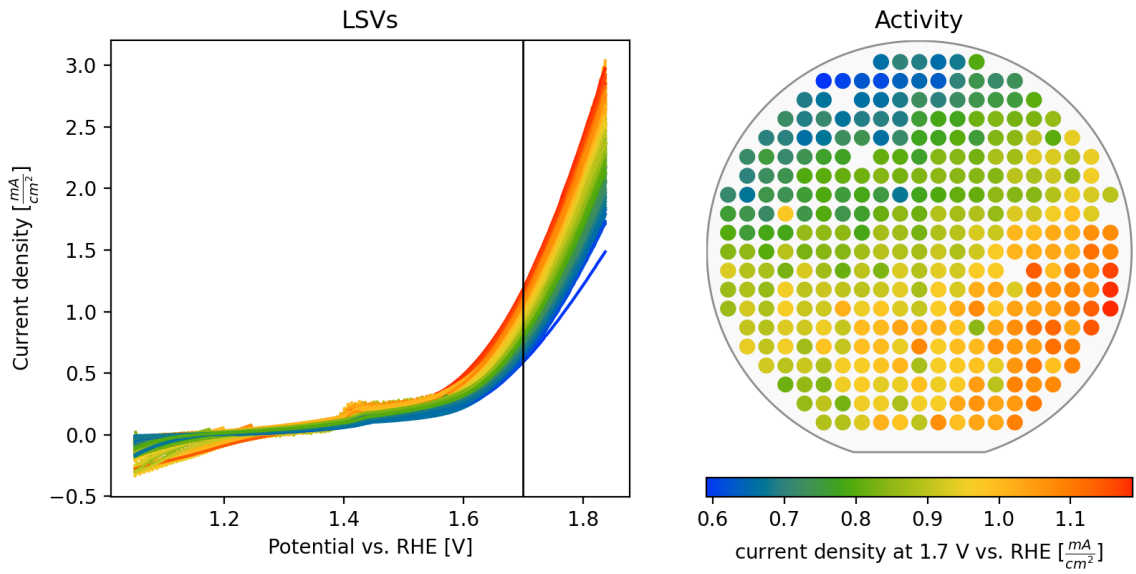

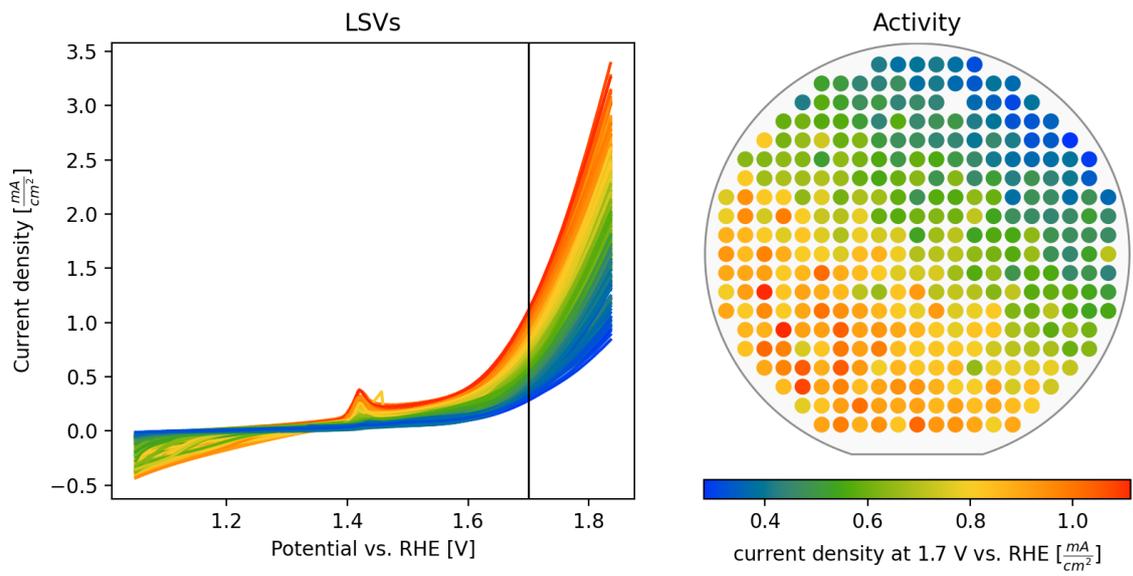